\documentclass[aps,prb,showpacs,twocolumn]{revtex4-1}

\usepackage{graphicx}

\begin{document}

\title{Dynamical properties of bidirectional charge-density-waves in ErTe$_3$}

\date{\today}

\author{A.A.~Sinchenko$^{1,2,3,4}$, P.~Lejay$^{2,3}$, O. Leynaud$^{2,3}$ and P.~Monceau$^{2,3}$}

\address{$^{1}$Kotel'nikov Institute of Radioengineering and Electronics of RAS, Mokhovaya 11-7, 125009 Moscow, Russia}

\address{$^{2}$University Grenoble Alpes, Institut Neel, F-38042 Grenoble, France}

\address{$^{3}$CNRS, Institut Neel, F-38042 Grenoble, France}

\address{$^{4}$National Research Nuclear University (MEPhI), 115409 Moscow,Russia}

\begin{abstract}

We report a strong difference in the sliding properties of the bidirectional charge density wave (CDW) in the two-dimensional rare earth tritelluride ErTe$_3$ which occurs below $T_{CDW1}=265$ K with a wave vector along $c$-axis and below $T_{CDW2}=165$ K with a wave vector along $a$-axis; the excess current carried by the motion of the CDW is 10 times less for the lower CDW compared with the value of the upper one. We tentatively explain this result by a stronger pinning of the lower temperature CDW intricated with the upper one, which inhibits its motion and may generate a phase slippage lattice.

\end{abstract}


\maketitle

\section{Introduction}\label{Intr}

Symmetry-breaking phase transitions with interplay between multiple degrees of freedom are intensively studied in strongly correlated systems such as high-$T_c$ superconductors, modulated charge or spin structures (charge/spin density waves) and another ordered states. A charge density wave (CDW) is characterized by a spatial periodic modulation $\sim \cos(Qx+\phi)$ of the electronic density concomitant with a lattice distortion with the same periodicity inducing the opening of a gap, $\Delta$, in the electronic dispersion. The collective properties of the CDW quantum state can be described by a complex order parameter $\Psi\sim\Delta e^{i\phi}$ with the collective excitations of $\Delta$ and $\phi$ being the amplitude and phase modes.

Although theoretically predicted by Kohn \cite{Kohn59}, Overhauser \cite{Overhauser62}, Peierls \cite{Peierls55} and Fr\"{o}hlich \cite{Frohlich54} in the 1950s and 1960s, the first Peierls-Fr\"{o}hlich transition was experimentally observed nearly simultaneously in the beginning of 1970s in quasi-one dimensional systems -- linear Pt-chain material \cite{Comes} [K$_2$Pt(CN)$_4$Br$_{0.30}$xH$_2$O], organic tetrathiafulvalene-–tetracyanoquinodimethane salt (TTF-–TCNQ)(see for review \cite{Heeger}), and quasi-two dimensional transition metal dichalcogenides MX$_2$ (M: Nb, Ta; X: S, Se)\cite{MX2}. Then a few years later, CDW transitions were found in transition metal trichalcogenides, NbSe$_3$, TaS$_3$\cite{Monceau12} and in molybdenium bronze K$_{0.3}$MoO$_3$\cite{bronze}.

In one hand, from 1D weak coupling mean field theories, with $\Delta/E_F\ll 1$, the Peierls instability is driven by the electronic energy gain which originates mostly from the Fermi surface nesting with $Q=2k_F$. In the other hand, in the strong limit coupling, $\Delta/E_F\simeq 1$, as recognized by McMillan\cite{McMillan77} and Varma and Simons\cite{Simons83} the transition is driven by the entropy of the lattice; the energy gain is then spread over the entire Brillouin zone as recently observed by inelastic neutron scattering\cite{NScatt}. As the electron phonon coupling is increased the importance of the electronic structure in $k$-space is reduced and a local chemical bonding picture in real space is more appropriate.

One of the most representative features of one-dimensional (1D)
compounds with a charge-density wave (CDW) is the possibility of
collective electron transport first predicted by H. Fr\"{o}hlich
\cite{Frohlich54}, as a model for superconductivity in 1D. The
extraconductivity which results from the CDW sliding is a collective
motion of electrons, a mechanism totally different from the
classical one of elastic scattering of individual electrons which
leads to the residual resistance. However various mechanisms such as
impurities, defects, interchain interaction or commensurability pin
the phase of the CDW and at low electric field the conductivity
exhibits a constant ohmic behavior due to quasi-particle excitations
only. To overcome the pinning energy and to initiate the CDW sliding
it is necessary to apply an electric field of a sufficient strength
larger than some characteristic threshold electric field, $E_t$
\cite{Gruner,Monceau12}. At the present time sliding CDW properties have
been observed and well studied in many inorganic as well as in
organic one-dimensional compounds (for recent review, see Ref.
\onlinecite{Monceau12}).

Since many years, the search of a possible CDW sliding in two dimensional (2D)compounds was unavailing. Only recently it was
succeeded in observing collective CDW motion in quasi-2D 
rare-earth tritellurides compounds \cite{SinchPRB12,SSC14}. This new family of
quasi-2D compounds has raised an intense research activity last time
\cite{DiMasi95,Brouet08,Ru08}, because it is considered as a
model system for which the structure of the CDW ground state can be
theoretically studied \cite{Yao06}. Thus a phase diagram as a
function of the electron-phonon parameter was derived with a
bidirectional (checkerboard) state if the CDW transition temperature
is sufficiently low whereas a unidirectional stripe state, as
observed experimentally, occurs when the transition temperature is
higher. This result is relevant for a deeper understanding of the
charge pattern in highly correlated materials, and particularly to
the recent determination of the biaxial CDW in underdoped cuprates
\cite{LeBoeuf2013}.

$R$Te$_{3}$ ($R$=Y, La, Ce, Nd, Sm, Gd, Tb, Ho, Dy, Er, Tm) layered
compounds have a weakly orthorhombic crystal structure (space group
$Cmcm$). They are formed of double layers of nominally square-planar
Te sheets, separated by corrugated $R$Te slabs. In this space group,
the long $b$ axis is perpendicular to the Te planes. These systems
exhibit an incommensurate CDW below the Peierls transition
temperature $T_{CDW1}$ through the whole $R$ series
\cite{Ru08,Lavagnini10R}, with a wave vector
$\mathbf{Q}_{CDW1}=(0,0,\sim 2/7c^{\ast})$ , with $T_{CDW1}$ above 300K for the light atoms (La, Ce, Nd). For the heavier $R$ (Tb, Dy, Ho, Er, Tm) atoms a second CDW
occurs at low temperature $T_{CDW2}$ with the wave vector
$\mathbf{Q}_{CDW2}=(\sim 2/7a^{\ast},0,0)$ perpendicular to
$\mathbf{Q}_{CDW1}$. The superlattice peaks measured from X-ray
diffraction are very sharp and indicate a long range 3D CDW order
\cite{Ru08}.

Below the Peierls transition, in all $R$Te$_3$ compounds, the Fermi
surface is partially gapped resulting in a metallic behavior at low
temperature. The layered $R$Te$_3$ compounds exhibit a large
anisotropy between the resistivity along the $b$-axis and that in
the $(a,c)$ plane, typically $\sim10^2$ below $T_{CDW1}$ and much
higher at low temperature \cite{Ru06}. Because the unidirectional
character of the upper CDW \cite{Fang07,Lavagnini10R,Yao06}, a
conductivity anisotropy in the $(a,c)$ plane arises in the CDW state
as was observed experimentally and explained theoretically in Ref.
\onlinecite{sinchPRL14}. The effect of the upper CDW on the in-plane
resistivity observed in experiments is very weak, no more than a few
percents of the total resistance \cite{Ru08,sinchPRL14,Ru06}.

Amplitude CDW excitations were probed by Raman spectroscopy\cite{Lavagnini10R}, time-resolved ARPES\cite{Leuenberger15} and femtosecond pump-probe spectroscopy\cite{Yusupov08,Moore16}. This later technique has allowed to study the disentanglement of the electronic and lattice path of the CDW order parameter, the collective vibrations being assigned to amplitude modes.
On the other hand, collective charge phase excitations could not be observed in far-infrared measurements due to screening by the residual metallic component of the Fermi surface. But the phase collective mode is accessible through non-linear transport properties.

Thus, effect of collective electron transport was observed for the high temperature CDW in single crystals of DyTe$_3$\cite{SinchPRB12}, TbTe$_3$ and GdTe$_3$\cite{SSC14} but only when the electric field is applied along the $Q_{CDW1}$-direction, namely $c$-axis. Sliding effect is completely absent when the current is applied in the perpendicular, $a$-axis direction, demonstrating the unidirectional character of the high-T CDW. No attempt up to now was made at observing the possible sliding of the low-T CDW in RTe$_3$.

Among all CDW compounds, only a few exhibit multiple CDWs. In MX$_2$ compounds, a triple $Q$ structure is formed with three wave vectors of equal amplitude, 120$^o$ apart\cite{MX2}. In TTF-–TCNQ a Peierls transition occurs first on TCNQ stacks at 54 K, a second CDW transition on parallel TTF stacks at 49 K which drives the transverse modulation along $a$ from $2a$ to a locked value $4a$ at 38 K\cite{Comes}. Both CDWs in NbSe$_3$ are formed on two different parallel chains\cite{Monceau12}. 

Hereafter we report measurements of CDW sliding properties of ErTe$_3$ which exhibit two CDWs at $T_{CDW1}=270$ K and $T_{CDW2}=165$ K. The situation is unique. Unlike NbSe$_3$ where both CDWs slide along parallel different chains, both CDWs in ErTe$_3$ exist within the same Te-planes, thus both CDWs modulate the positions of the same Te atoms. The main questions we would like to answer are the following:

(i) is it possible to detect the sliding of the low-T CDW, besides the fact that the anomaly of resistivity at $T_{CDW2}$ is barely visible? 

(ii) is there any change of the sliding of the high-T CDW below $T_{CDW2}$? 

(iii) are these orthogonal CDWs totally independent or are they
interacting one with the other? If yes, what is the result and the
mechanism of this interaction?

\section{Experimental}

Single crystals of ErTe$_3$ were grown by a self-flux technique
under purified argon atmosphere as described previously
\cite{SinchPRB12}. For experiments we chose good quality single
crystals. Thin single-crystal samples with a thickness less than 2
$\mu$m were prepared by micromechanical exfoliation of relatively
thick crystals glued on a sapphire substrate. The quality of
selected crystals and the spatial arrangement of crystallographic
axis were controlled using X-ray diffraction techniques. X-ray
diffraction measurements show that single crystals with a thickness
more than 2 $\mu$m as a rule are twinned with the change between $c$
and $a$ axis in neighboring layers. However, for thinner samples it
was possible to select untwinned single crystals with well defined
crystallographic axis positions. 

From untwinned single crystals with a thickness typically 0.2-2.0
$\mu$m, we cut stripes with a length of a few mm and a width $50-80$ $\mu$m in well defined orientation, namely [100] and [001]. In the following we call $c$-axis stripes samples with the length along the [001] direction and $a$-axis stripes those with the length along [100]. In some cases it was succeeded
to select native crystals with a shape of a narrow stripe with a width
50-100 $\mu$m and length near 1 mm, oriented along $c$-axis. Measurements of current-voltage characteristics (IVs) and their
derivatives have been performed with a conventional 4-probe
configuration. For contacts preparation we used gold evaporation and
cold soldering by In. The current was applied along the length of the stripes. For studying nonstationary effects a radiofrequency (rf) current was superposed on the dc current using current contacts connected with the generator via two capacitors.
All measurements have been performed in the temperature range
4.2-340 K. 

Example of temperature dependencies of the normalized resistance, $R(T)/R(300K)$, for $c$- and $a$-axis stripes and conductivity anisotropy, $R_a(T)/R_c(T)$ is shown in Fig.\ref{F1}. The difference between the T-dependence of $R_a$ and $R_c$ is a clear indication that the sample is untwinned. The dependencies shown in Fig.\ref{F1} corresponds qualitatively to those reported in Ref. \onlinecite{sinchPRL14} for DyTe$_3$, HoTe$_3$ and TbTe$_3$. However, one can note the decrease of anisotropy $R_a/R_c$ below 40 K the origin of which needs further investigations. 

\section{Experimental results}

\subsection{high-T CDW}

As reported before, the collective CDW motion for the high-T CDW is observed only in $c$-axis stripes. We were able to measure characteristic nonlinearity in IV curves below $T_{CDW2}$ down to 100 K; the conductivity of the stripe increases sharply above a threshold electric field. However, the shape of IV characteristics change qualitatively for temperatures slightly above and below $T_{CDW2}$ as demonstrated in Fig. \ref{F2}.
As can be seen, at temperatures below $T_{CDW2}$ IV curves become
hysteretic and much more noisy. Additionally, sharp maxima of
$dV/dI$ appear at electric field close to $E_t$.

The temperature evolution of differential IV curves in the
temperature range 270-110 K for one of the stripes is shown in
Figure \ref{F3}. Neighbour curves differ by $\Delta T=5$ K. The
same behavior was observed for another stripes. Note that the relative change of the differential resistance from static to sliding state is very small, no
more than 3\% from the total resistance, indicating a very low
contribution of the CDW to the electron transport.

It can be seen as drawn in Fig.\ref{F8} that the threshold electric field increases
monotonically with the decrease of temperature but in the temperature
range $T\approx 190-160$ K $E_t$ increases more rapidly. Nearly the same effect was observed in the quasi-one dimensional NbSe$_3$ for high-T CDW at temperatures close to the second CDW transition \cite{Q1Q2}. Note that in these $c$-axis stripes (with the current applied along the $c$-axis) below $T_{CDW2}$ there are no indication of sliding of the low-T CDW the $Q$-vector of which being along $a$-axis.

CDW sliding is accompanied by low-frequency broadband electric
noise (BBN) and narrowband noise (NBN). The collective electronic
transport and the NBN generation can be characterized in terms of
the CDW coherence and homogeneity of the CDW in space and in time.
The coherence can be affected by external rf (or hf) irradiation.
The most widely studied effect of the irradiation is the
synchronization of the CDW, known also as the interference effect,
mode locking, or Shapiro steps.\cite{Monceau12}. In the present
study we have observed Shapiro steps at such experimental
conditions. Fig.\ref{F4} shows $dV/dI(I)$ dependencies at $T=230$
K under application of a rf field with different frequencies from 2
MHz up to 11 MHz with the rf ac field amplitude of 900 mV. Shapiro steps are
clearly observed in the $dV/dI(I)$ characteristics at all frequencies
as sharp maxima in the differential resistance demonstrating high
level coherency of the CDW. 

It is interesting also to trace the evolution of Shapiro steps with temperature at fixed
frequency. Fig. \ref{F5} shows $dV/dI(I)$ curves
under application of a rf field with $F=4.5$ MHz and amplitude $V=0.9$ V at different
temperatures. As can be seen, Shapiro steps are clearly observed
only down to temperatures close to $T_{CDW2}$. Close to this temperature
the amplitude of Shapiro steps starts to decrease that indicates the loss of
coherency of the high-T CDW in the temperature range where the low CDW occurs.

As shown in Fig.\ref{F4} the separation $\Delta I$ between Shapiro steps increases when the rf-frequency increases. Having two contributions to the electric current: normal
electrons and collective CDW transport, $\Delta I_{CDW}$ can be easy
calculated using dc IV characteristics:

\begin{equation}
\Delta I_{CDW}=I_{total}(1-\frac{R}{R_{N}})
\label{deltaI}
\end{equation}

where $I_{total}$ - total current, $R$ - actual resistance of the sample
and $R_N$ - normal state resistance.

In Fig.\ref{F9} we have plotted the excess current density $\Delta J_{CDW}=\Delta I_{CDW}/S$ with the cross-section of the sample $S=2.1\times10^{-6}$ cm$^{-2}$ as a function of rf frequency. As can be seen, as in quasi-1D systems\cite{Monceau12}, $\Delta J_{CDW}$ linearly increases with frequency.

\subsection{low-T CDW}

In $a$-axis stripes prepared from untwinned single-crystals we
succeeded to observe collective motion for the low-T
CDW. A typical IV curve demonstrating characteristic CDW non-linearity
at $T=140$ K well below $T_{CDW2}$  is shown
in Fig. \ref{F6}(a). As can be seen, at $E>E_t=0.44$ V/cm, the
differential resistance sharply decreases.

The temperature evolution of differential IV curves in the
temperature range 90-170 K for the same sample is shown in Figure
\ref{F6}(b). The curves are shifted for clarity and the difference between each of them is $\Delta T=5$ K. As can be seen, the threshold electric field weakly decreases with the increase of $T$ in the range 90-150 K. In contrast to the high-T CDW, at
temperatures close to $T_{CDW2}\approx165$ K $E_t$ starts to increase
indicating divergency at $T_{CDW2}$. Such a behavior has been previously observed in
quasi-1D compounds with a CDW \cite{Monceau12}. The T-dependence of the threshold field $E_t$ for the low-T CDW is plotted in Fig.\ref{F8}. There are no any
non-linearity in the IV-characteristics at temperatures above $T_{CDW2}=165$ K
except a very weak Joule heating. Note that the observed contribution to
the electron transport from the low-T CDW sliding is nearly 4 times
less compared with the high-T CDW and its amplitude is no more than 0.7\%
from the total current.

To confirm that the observed non-linearity of IV curves is the real
sliding of the low-T CDW we measured IV characteristics under application
of dc and rf electric field. As in the case of the high-T CDW, in
spite of the low amplitude in the sliding effect, we observed
pronounced Shapiro steps however. Figure \ref{F7}(a) shows
$dV/dI(I)$ dependencies at $T=130$ K under application of a rf field
with a frequency of 28 MHz and amplitude 0.9 V. For comparison, the static (without rf
field) differential IV curve (blue) measured at this temperature is also
shown. Note that the application of a rf electric field leads to a
reduction of the threshold electric field $E_t$. At the same time,
Shapiro steps appear in the $dV/dI(I)$ characteristics as sharp maxima
in the differential resistance. With increasing frequency the
distance between neighboring maxima increases proportionally to the
frequency; that is illustrated in Fig. \ref{F7}(b) where we show IV
curves under application of a rf field with a frequency of 7, 14,
21, 35 and 55 MHz at the same temperature and with the same rf
power. The curves are shifted relative to each over for clarity. The excess current density calculated from Eq.\ref{deltaI} (with the cross-section of the sample $S=0.25\times10^{-6}$ cm$^{-2}$) is plotted in Fig.\ref{F9} as a function of rf frequency. As for the upper CDW $\Delta J_{CDW}$ linearly increases with frequency.

\section{Discussion}

The RTe$_3$ family (R: La, Ce, Pr, Nd, Gd, Tb, Dy, Er, Tm) is a model system for studying the effect of chemical pressure on the Peierls transition due to a reduction of the in-plane lattice constant from light R (La, Ce, ...) towards heavier R ions (Dy, Er, Tm). While the $\mathbf Q_{CDW1}$ wave vector along $c$-axis (which results from charge transfer between RTe buckled planes and Te planes) is nearly the same for each member of the family, the Peierls transition temperature is much larger than room temperature for light R and decreases below room temperature for heavier R \cite{Ru08}. From optical spectroscopy \cite{Sachetti09} and ARPES measurements \cite{Brouet08} it was shown that the CDW gap scales with the lattice parameters. In addition, optical measurements \cite{Pfuner10} have revealed that the remaining fraction of ungapped FS in the CDW state is larger for compounds with smaller lattice parameters. Thus phenomenologically, it was suggested that a second CDW is formed only when the first CDW is weakened with the decrease of the lattice parameter, making larger FS sections available for the new nesting condition in the transverse $a$-axis.

While ARPES measurements \cite{Brouet08}are interpreted as a strong evidence for a FS nesting scenario, inelastic x-ray scattering  \cite{Maschek15} and Raman experiments \cite{Eiter13} emphasize the strongly momentum-dependent electron-phonon coupling. In that case, a local chemical description of the distortions is more realistic, as proposed in Ref.\onlinecite{Malliakas06} where the distortion of the Te net is viewed as an oligomer sequence of Te trimers and tetramers. 

As far as ErTe$_3$ is concerned, only a few publications are available. The low-T CDW was discovered by Ru et al.\cite{Ru08}.The superstructures $Q_{CDW1}=0.298 c^*$ and $Q_{CDW2}=0.313 a^*$ (measured at $T=10$ K) are sharp. Both wave vectors are present in the same crystallite. The integrated intensity of $Q_{CDW2}$ exhibit large fluctuations above $T_{CDW2}$, at least up to 180 K. The CDW gaps were determined by ARPES \cite{Moore10} with $\Delta_{CDW1}=175$ meV and $\Delta_{CDW2}=50$ meV with the temperature dependence of $\Delta_{CDW1}$ slightly suppressed from the mean-field variation. These values are in agreement with those measured by Raman scattering \cite{Eiter13}. The ratio $2\Delta_{CDW1}/k_BT_{CDW1}\simeq15$ is much larger than the BCS mean field 3.52 value, similarly to many one-dimensional systems\cite{Monceau12} indicating strong coupling effect or the role of fluctuations.The observation of the amplitude mode by Raman scattering \cite{Eiter13} such $\omega_{AM}=\sqrt{\lambda}\omega_{2k_F}$ with $\omega_{2kF}$ the frequency of the unnormalized CDW phonon energy yields $\lambda=0.4$ indicating strong coupling.

It has to be noted \cite{Moore10} that the lower CDW is weaker with
\begin{equation}
\frac{2\Delta_{CDW1}}{k_BT_{CDW1}}\sim 2\frac{2\Delta_{CDW2}}{k_BT_{CDW2}}
\label{DeltaCDWs}
\end{equation}

The density of states was calculated from the interacting tight-binding model. The onset of the high-T CDW suppresses $N(E_F)$ by $\sim$ 77\% of the unmodulated value while the low-T CDW further suppresses $N(E_F)$ by $\sim$ 74\% revealing that the gain in the second CDW is really modest \cite{Moore10}.

Many models for the RTe$_3$ electronic structure consider only Te planes, sometimes a single one; the buckled Te slabs are viewed as a simple charge reservoir that determines the Fermi level with the f states of the rare earth localized away from the Fermi surface and expected to play no role. However the hybridization between the rare-earth 4f electrons and Te 2p electrons was revealed \cite{Lee12} by the observation of a diffraction peak near the M$_5$(3d-4f) absorption edge of rare-earth ions with a wave vector identical to that of the CDW.

We have experimentally measured nonlinear transport properties in ErTe$_3$ associated to each CDW, the upper one with current $I$ applied parallel to $Q_{CDW1}$ along $c$-axis, and the lower one with $I\parallel Q_{CDW2}$ along $a$-axis. Are these nonlinearities the signature of the Fr\"{o}hlich-type conductivity as demonstrated \cite{Monceau12} in quasi one-dimensional compounds?

As was mentioned in section \ref{Intr}, it is assumed that both CDWs exist within the Te planes and that both CDWs modulate the position of the same Te atoms. The possibility that CDWs occur on a different Te plane of the Te bilayer is very unlikely, although evoked in Ref.19. However it is worth to note that by Raman scattering experiments, the amplitude mode in the high-T CDW develops as a succession of two mean field transitions with different critical temperatures; that was associated to the Te bilayers \cite{Lavagnini10R}.

The main results we have obtained can be summarized as follows: in cooling, when approaching the bidirectional CDW ground state at $T_{CDW2}\simeq165$ K, the threshold field of the high-T CDW increases more than linearly, noise appears in the differential IV characteristics and the high-T CDW coherence is lost by the disappearance of Shapiro steps. In Fig. \ref{F8} we have plotted the temperature dependence of the 
threshold electric field for both CDWs. Taking into account the
fact that the absolute value of $E_t$ is sample dependent and
that the threshold characteristics of high- and low-T CDWs were
measured in different stripes, exfoliated however from the same single crystal, the curves in Fig. \ref{F8} demonstrate only the qualitative behavior of $E_t$.

As can be seen, in the temperature range 270-200 K the dependence $E_t(T)$ of the upper CDW is linear and can be well described by the expression:

\begin{equation}
E(T)=E(0)(1-\frac{T}{T_0}) \label{lin}
\end{equation}

with $E(0)=235$ mV/cm and $T_0=1.07T_{CDW1}$ similarly to the behavior of $E_t$ for the upper CDW reported previously \cite{SSC14}. In the range
200-165 K we observe a deviation from the linear dependence and $E_t$
increases more rapidly in this temperature range. At temperature
$T<T_{CDW2}=165$ K, $E_t(T)$ resumes again a linear dependence. We understand such a behavior as the result of interaction between high-T and low-T CDWs. Indeed it has been shown that fluctuations of the low-T CDW extend up to nearly 200 K \cite{Ru08}.

The depinning process for the unidirectional CDW, namely in TbTe$_3$ at room temperature, has been probed by coherent x-ray diffraction \cite{Bolloch16}.Contrary to one-dimensional systems (such as NbSe$_3$ and K$_{0.3}$MoO$_3$)\cite{Monceau12} the CDW remains undeformed below threshold and suddenly rotates to overcome pinning centers and reorders by motion above threshold.

In a superfluid there are a macroscopic occupation of a quantum state that picks out a unique reference frame which describes the velocity $v_s$ of the superfluid. In a superconductor it is the common momentum of Cooper pairs that defines $v_s$. In the Fr\"{o}hlich model, as stated by Allender et al.\cite{Allender74}, $v_s$ is determined by the velocity of the macroscopically occupied lattice wave which produces energy gaps. Then the extracurrent carried by the CDW into motion is $j=nev$. The frequency of Shapiro steps in differential IV characteristic was identified as the signature of the CDW velocity with $v=\lambda f$ with $\lambda$: the CDW wavelength. In Fig. \ref{F9} we have plotted together the excess current density in the nonlinear state of ErTe$_3$ as a function of the frequency of Shapiro steps for the high-T CDW at $T=230$ K and the low-T CDW at $T=130$ K. For the high-T CDW, in Fig. \ref{F9} we can evaluate the ratio J$_{CDW1} $ / {f$_0$ = 28 A/MHz cm$^2$. This value is very similar to the average value obtained on 14 samples for the upper CDW on NbSe$_3$: 40.2 A/MHz cm$^2$ and 24 A/MHz cm$^2$ for the lower one \cite{Richard93}. Similar values were also obtained for o-TaS$_3$ \cite{Zettle83,Sokoloff90}.

In a one-band model (or for one chain in 1D description), $\lambda=2\pi/2k_F$ and the condensate density is $2k_F/\pi$. Then
\begin{equation}
\frac{J_{CDW}}{f_0}=2e 
\label{j}
\end{equation}
This relation is well satisfied in 1D systems\cite{Monceau12} and demonstrates the Fr\"{o}hlich-type of conductivity.

Determining a similar value of $J_{CDW}/f_0$ for the high-T CDW in ErTe$_3$ indicates that the same Fr\"{o}hlich process is operating. Let consider the number of unit cells in the cross section of our sample with the lattice parameters $b=25.02 \mathring{A}$ and $a=4.29 \mathring{A}$, one get $1.96\times10^8$ unit cells. Consider also that along the $c$-axis, one have one Te chain per unit cell. Then we recover the value of $J_{CDW1}/f_0=28$ A/MHz cm$^2$.

We can also estimate the CDW velocity. At $T=240$ K, at the total current applied $I=10$ mA, we have evaluated $I_{CDW}=0.27$ mA. With the electron density $\sim10^{23}$ cm$^{-3}$ and taking into account that no more than 30\%-40\% of the FS is affected by the formation of the CDW \cite{Lavagnini10R}, we get a very small value of the CDW velocity $\sim 10^{-2}$ cm/s. However the number of carriers condensed below the CDW gap needs to be verified.

While the sliding properties of the high-T CDW appear to correspond to the Fr\"{o}hlich mechanism, the situation is totally different for the low-T CDW. From Fig. \ref{F9} we deduce the slope of $J_{CDW2}/f_0$ 10 times lower than the value for the high-T CDW. Additionally the threshold field is higher. This result is opposite to NbSe$_3$ with both CDWs sliding along parallel chains and for which the $J_{CDW2}/f_0$ is nearly the same for both CDWs \cite{Richard93}.

We can only speculate on a possible explanation of these results. The bidirectional CDW is formed by two orthogonal modulations of Te atoms which naturally interact and are imbricated. Because the sliding is along $c$-axis for the high-T CDW and along $a$-axis for the low-T CDW, the coupling between them, say bonds, should be broken. The crystallographic structure should have an important role. It was shown \cite{Ru08} that the formation of the high-T CDW (in TbTe$_3$) appears to "stretch" the lattice from its expected value along the direction of the modulation wave vector ($c$-axis). It is also the direction of the glide between the two Te planes. Then the depinning along $c$-axis may appear to be easier, even if the high-T CDW looses its coherence when crossing $T_{CDW2}$. 

From the general Fr\"{o}hlich mechanism, with the current carried by the motion of the CDW as $J_{CDW}=nev$ and with $v$ identified from Shapiro steps as $v=\lambda f_0$, then $J_{CDW}/f_0=ne\lambda$. In ErTe$_3$ the slop $J_{CDW}/f_0$ is 10 times less for the lower CDW with respect to the upper one. One may then suggest that the full electronic density condensed below the gap in the band of low CDW, as in the Fr\"{o}hlich model, does not participate to the conductivity but only a part of it, namely arround $1/10$. Pinning along $a$-axis may be stronger and the bond between both modulations anisotropic. Nonlinearity may result from the motion of a lattice of phase defects which is formed in the checkerboard lattice of both CDWs.   

This interpretation needs naturally to face some experimental results. It appears to determine the modulated structure of ErTe$_3$ in the unidirectional CDW state as well as in the bidirectional one. STM measurements at low temperature are also crucial as well as a theoretical model for sliding of a bidirectional CDW. However, STM images may be difficult to be interpreted because possibly blurred by the disorder \cite{Robertson06} of the bidirectional superstructure.

In conclusion we have observed the nonlinearity in transport properties of the bidirectional charge density wave ground state of ErTe$_3$. While the sliding properties of the upper CDW appear to be similar to those previously found in quasi one-dimensional systems, the nonlinearity for the lower CDW may involve a phase defect lattice. More works are naturally needed to ascertain or weaken the present interpretation.

\acknowledgements

The work has been supported by Russian State
Fund for the Basic Research (No. 14-02-01126--a), and partially
performed in the frame of the CNRS-RAS Associated International
Laboratory between CRTBT and IRE "Physical properties of coherent
electronic states in coherent matter".

\newpage

\begin{figure}[t]
		\includegraphics[width=8cm]{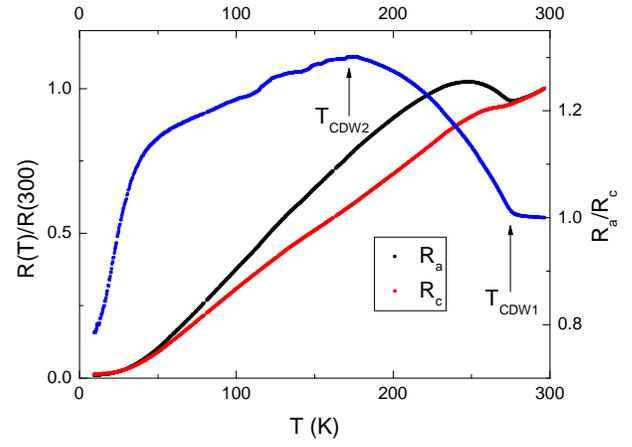}
		\caption{(color online) ErTe$_3$: temperature dependencies of normalized
		resistance, $R(T)/R(300K)$, and conductivity anisotropy,
		$R_a(T)/R_c(T)$, for $c$ and $a$-axis stripes} \label{F1}
\end{figure}

\begin{figure}[t]
		\includegraphics[width=8cm]{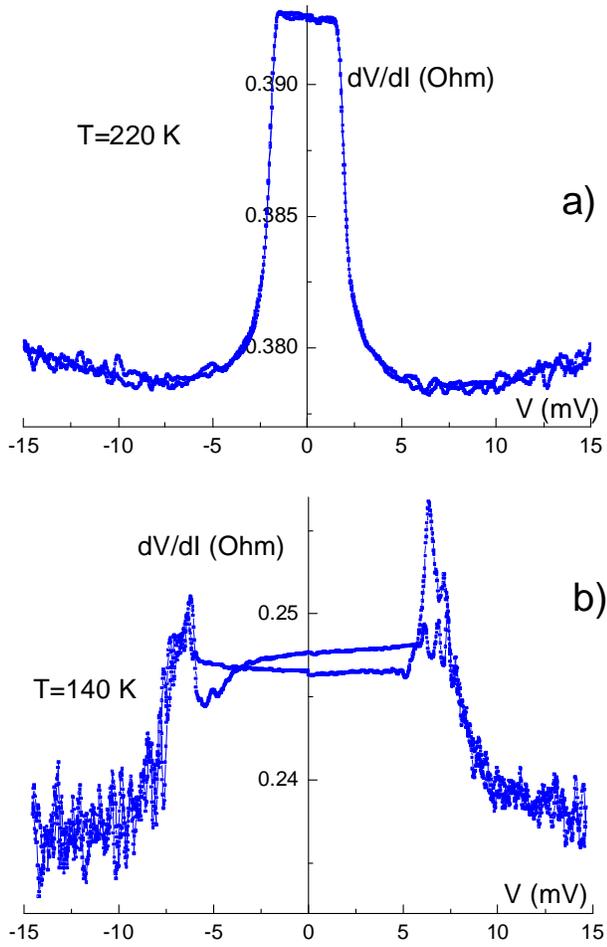}
		\caption{(color online) Differential current-voltage
		characteristics, $dV/dI(E)$ of a ErTe$_3$ stripe oriented along
		$c$-axis below $T_{CDW1}$ at $T=220$ K (a) and below $T_{CDW2}$ at $T=140$ K (b)} \label{F2}
\end{figure}

\begin{figure}[t]
		\includegraphics[width=8cm]{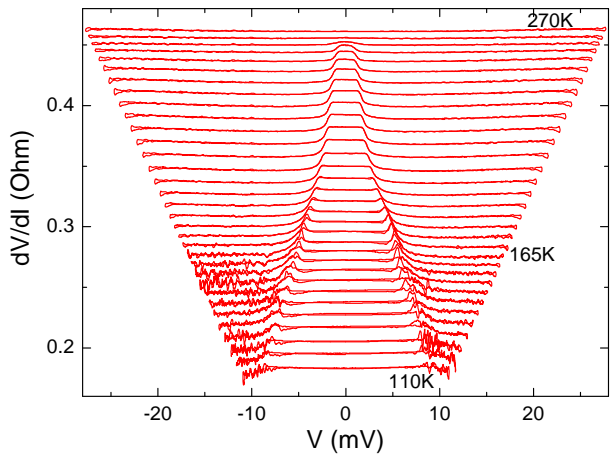}
		\caption{(color online) Temperature evolution of differential IV
		curves in the temperature range 270-110 K for one of the ErTe$_3$ stripes
		oriented along $c$-axis.} \label{F3}
\end{figure}

\begin{figure}[t]
		\includegraphics[width=8cm]{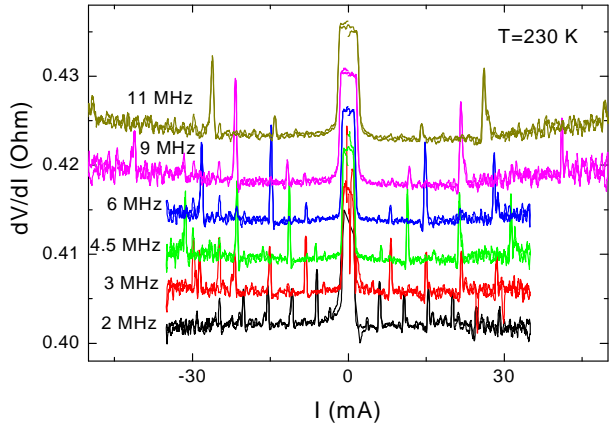}
		\caption{(color online) $dV/dI(I)$ dependencies of ErTe$_3$ at $T=230$ K
		under application of a rf field with different frequencies from 2
		MHz up to 11 MHz with the rf ac field amplitude of 900 mV for the same sample
		as shown in Fig.\ref{F2}. The curves are shifted for clarity.}
	\label{F4}
\end{figure}

\begin{figure}[t]
		\includegraphics[width=8cm]{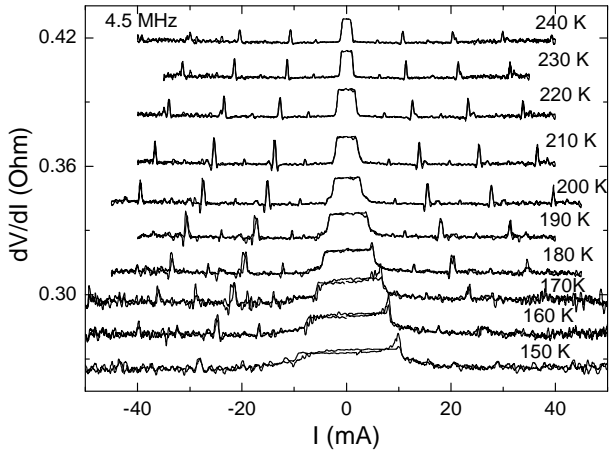}
		\caption{ErTe$_3$: $dV/dI(I)$ dependencies under application
		of a rf field with frequency 4.5 MHz at different temperatures for a ErTe$_3$ stripe oriented along $c$-axis.} 
	\label{F5}
\end{figure}

\begin{figure}[t]
		\includegraphics[width=8cm]{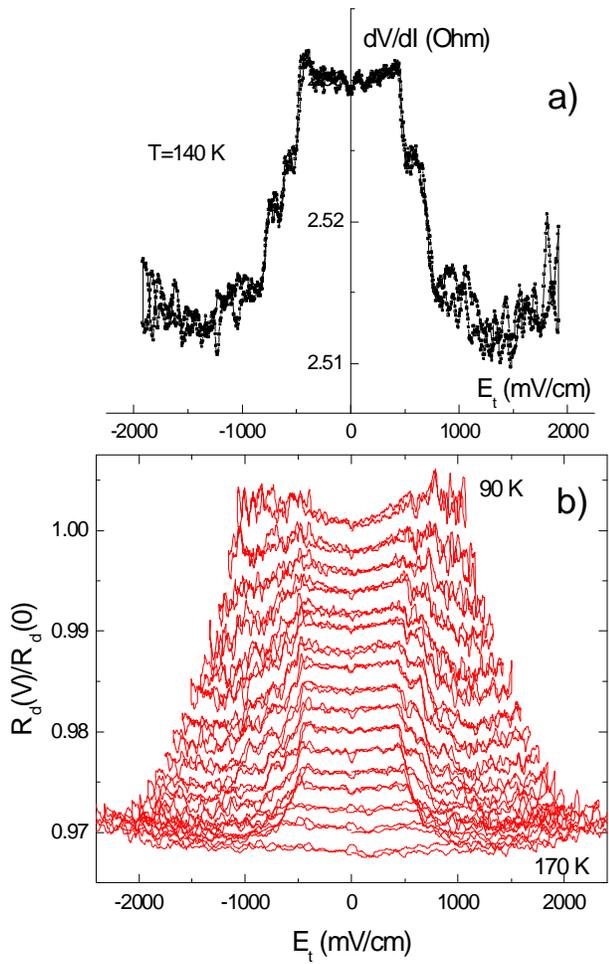}
		\caption{(color online) a) Differential current-voltage
		characteristics, $dV/dI(E)$ of a ErTe$_3$ stripe oriented along
		$a$-axis at $T=140$ K. b) Temperature evolution of normalized differential IV
		curves in the temperature range 90-170 K; the difference between each curve is 5 K.}
	\label{F6}
\end{figure}

\begin{figure}[t]
		\includegraphics[width=8cm]{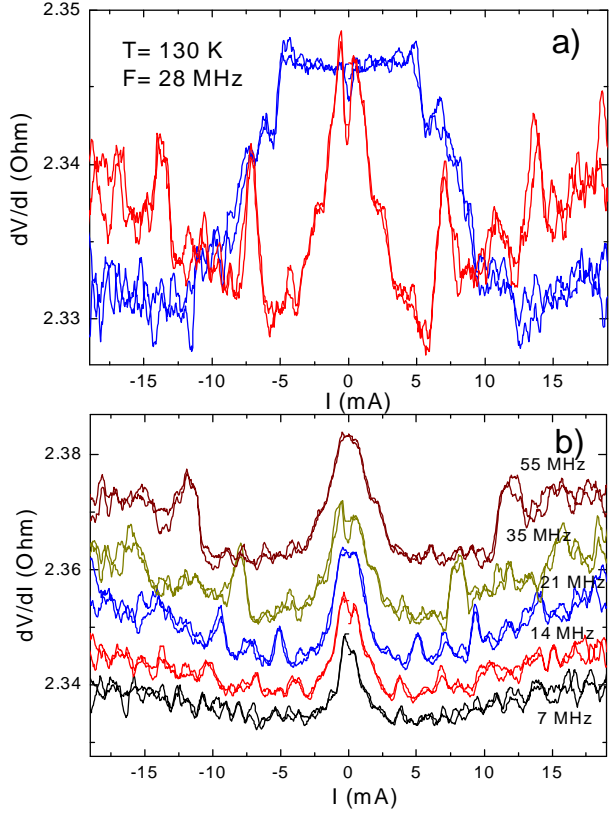}
		\caption{(color online) a) $dV/dI(I)$ dependencies at $T=130$ K under
		application of a rf field with a frequency of 28 MHz (red) and without rf
		electric field (blue) for a ErTe$_3$ sample oriented along $a$-axis. b) $dV/dI(I)$
		dependencies at $T=130$ K under application of a rf field with a
		frequency of 7, 14, 21, 35 and 55 MHz at the same temperature and
		with rf power 0.9V. The curves are shifted for clarity.} 
	\label{F7}
\end{figure}

\begin{figure}[t]
		\includegraphics[width=8cm]{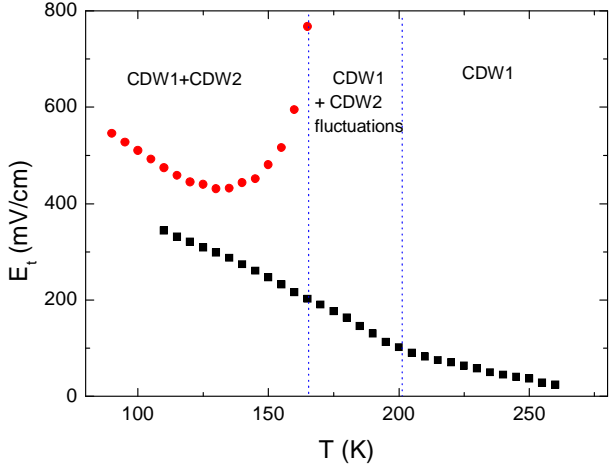}
		\caption{(color online) Temperature dependence of the threshold
		field $E_{T}$, for high- and low-T CDW in ErTe$_3$.} \label{F8}
\end{figure}

\begin{figure}[t]
		\includegraphics[width=8cm]{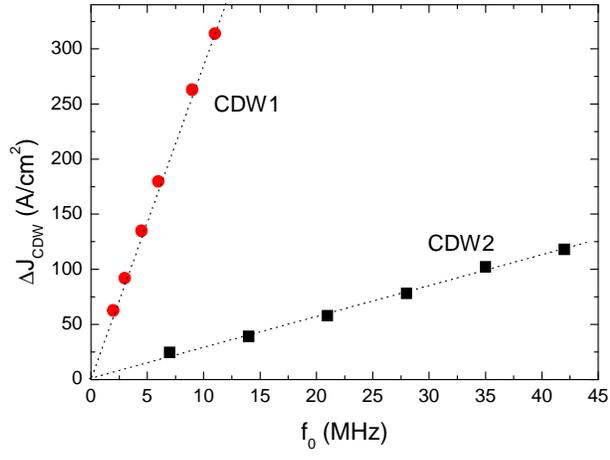}
	\caption{(color online) Excess current density in the nonlinear state of ErTe$_3$ as a function of the frequency of Shapiro steps for the high-T CDW at $T=230$ K and the low-T CDW at $T=130$ K.} \label{F9}
\end{figure}

\end{document}